\documentclass{PoS}
\usepackage{cite}

\title{The future SIDIS measurement on 
transversely polarized deuterons by the COMPASS Collaboration}

\ShortTitle{d-quark transversity}

\author{\speaker{Franco Bradamante}\\
  INFN, Trieste , Italy\\
  E-mail: Franco.Bradamante@ts.infn.it}

\author{on behalf of the COMPASS Collaboration}
\abstract{
Since 2005, measurements of Collins and Sivers asymmetries from the HERMES and COMPASS experiments
have allowed to assess that the transversity and the Sivers PDFs are different from zero and
measurable in semi-inclusive DIS on transversely polarised targets. Most of the data were collected
on proton targets, only small event samples were collected in the early phase of the COMPASS
experiment on a deuteron ($^6$LiD) target and more recently at JLab, on $^3$He, so that the
$d$-quark and the sea-quarks PDFs are much more poorly known than the $u$-quark PDFs. This
constitutes an important limitation to the knowledge of the transverse spin structure of the
nucleon. For this reason the COMPASS Collaboration has proposed to measure semi-inclusive DIS on
transversely polarised deuterons with good accuracy, comparable with that of the existing transverse
spin asymmetry data on protons. The proposal has been accepted by CERN and the experiment will run
in 2021, as soon as the Long Shut-down 2 is over, providing measurements which will stay unique for
many years to come. Projections will be given for the extraction of the transversity PDFs, and for
the evaluation of the isovector tensor charge of the nucleon. }

\FullConference{
23$^{rd}$ International Symposium on Spin Physics\\
Spin 2018 Ferrara, Italy\\
September 10-14, 2018}

\begin{document}

\section{Introduction}

The description of the nucleon structure in hard inclusive processes in terms of collinear parton
distribution functions (PDFs) has recently been generalised to take into account the transverse
momentum $k_T$ of the parton with respect to the nucleon direction. At leading twist a complete
picture of the nucleon requires a total of eight transverse-momentum-dependent (TMD) distributions,
which depend on the parton light-cone momentum fraction $x$, on a characteristic hard scale $Q^2$,
and on $k_T$ (for reviews, see~\cite{Barone:2010zz,Aidala:2012mv,Avakian:2016rst}).

When the target nucleon is transversely polarised, eight different spin-dependent azimuthal
modulations are expected in the SIDIS cross-section, so that from a measurement of the cross-section
and knowing the fragmentation functions (FFs) important information on the TMD PDFs can be
extracted. In this domain the HERMES~\cite{Airapetian:2010ds,Airapetian:2009ti} and the
COMPASS~\cite{ Ageev:2006da,Alekseev:2010rw,Adolph:2012sn,Adolph:2012sp} Collaborations have
performed pioneering measurements at different beam energies (27 and 160 GeV$/c$ respectively) and
shown beyond any doubt that two spin-dependent azimuthal modulations of the produced hadrons, of the
type $ \left[1 + A_C \sin( \phi_h + \phi_S - \pi)\right] $ and $ \left[1 + A_S \sin( \phi_h -
 \phi_S)\right] $, are clearly different from zero. The azimuthal angles of the produced hadron
transverse momentum and of the spin direction of the target nucleon are defined with respect to the
lepton scattering plane, in a reference system in which the z axis is the virtual-photon
direction. The first modulation, called Collins asymmetry, is described as a convolution of the
transversity function and a newly introduced TMD FF, the Collins FF, $H_{1}^{\perp}$. The second
modulation, called the Sivers asymmetry, is described as a convolution between the Sivers function,
a new TMD PDF, and the usual fragmentation function. For all details I refer again to the existing
literature ~\cite{Barone:2010zz,Aidala:2012mv,Avakian:2016rst}). The fact that the Collins
asymmetry and the Sivers asymmetry have been shown to be non zero (at least on a proton target)
implies that:
\begin{itemize}
\item[-] the Collins FF $H_{1}^{\perp}$ is different from zero
\item[-] the transversity PDF, $h_1$, is different from zero
\item[-] the Sivers PDF, $f_{1T}^{\perp}$, is different from zero.
\end{itemize}

\begin{figure}[b]
\begin{center}
\includegraphics[width=0.9\textwidth]{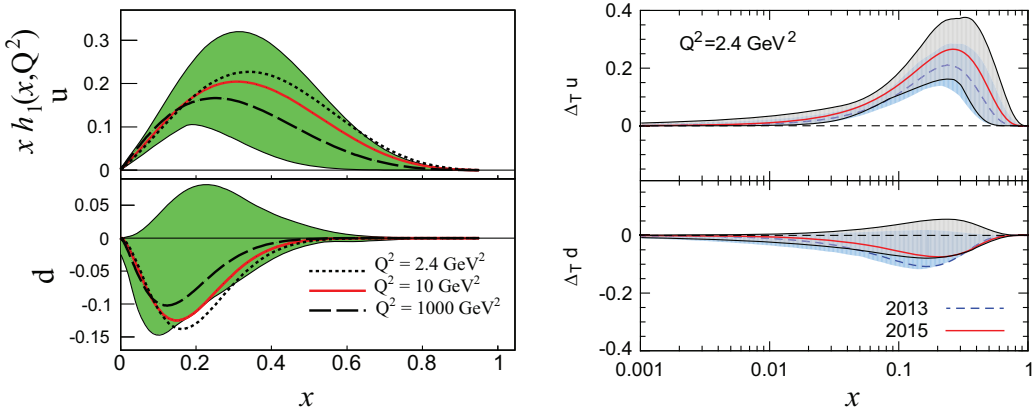}
\end{center}
\vspace{-0.5cm}
\caption{The $u$ and $d$ quark transversity PDFs from recent global fits. The plots are from
 Ref.~\cite{kang:2015msa} (left) and from Ref.~\cite{Anselmino:2015sxa} (right). Although the
 notation for the transversity PDF in the plot at the right is not the one used in this paper, the
 same function is shown in both plots.}
\label{fig:extr2}
\end{figure}

Independent evidence that the Collins FF is different from zero was provided by the measurement of
azimuthal asymmetries in inclusive production of hadron pairs in $e^+ e^-$ annihilation by the Belle
Collaboration~\cite{Seidl:2008xc}. When combining the HERMES~\cite{Airapetian:2010ds} and the
COMPASS~\cite{ Ageev:2006da,Alekseev:2010rw,Adolph:2012sn} data with the Belle measurements of $e^+
e^- \rightarrow hadrons$, first extractions of the Collins FF and of the transversity PDF have been
possible. Some recent extractions of the transversity function, from Ref.~\cite{kang:2015msa} and
from Ref.~\cite{Anselmino:2015sxa}, are shown in Fig.~\ref{fig:extr2}. In spite of the large
uncertainties, it is an important achievement. The trend of the functions is similar to that of the
$u$ and $d$ helicity distributions, but clearly we are still a long way from being able to assess
the difference between the transversity and the helicity PDFs, which is the ultimate goal of this
investigation.
A second feature in Fig.~\ref{fig:extr2} is that the uncertainty in $h_1^q$ for the $d$ quark is
much larger than in $h_1^q$ for the $u$ quark. This is due to the fact that most of the existing
data have been collected on proton targets, and only few measurements exist on neutron targets.
COMPASS has measured transverse spin asymmetries using both proton and deuteron targets. On the
deuteron, the asymmetries are compatible with zero~\cite{ Ageev:2006da}, but the accuracy of the
measurements is definitely inferior to that of the proton data. More recently data have been
collected at much lower energy at JLab on a $^3$He target, essentially a transversely polarised
neutron target: the measured asymmetries~\cite{Qian:2011py,Zhao:2014qvx} are also compatible with
zero, but the error bars are again fairly large.

The measurement of the transversity distributions is particularly important because they provide
access to the tensor charges $\delta q$, or $g_T^q$, which are given by the integral
\begin{eqnarray}
\delta q (Q^2)&=&\int_0^1 dx [h_1^q(x,Q^2)-h_1^{\bar{q}}(x,Q^2)]= \int_0^1 dx [h_1^{q_v}(x,Q^2)]
\end{eqnarray}
In a non-relativistic quark model, $h_1^q$ is equal to $g_1^q$, the helicity distribution, and
$\delta q$ is equal to the valence quark ${q_v}$ contribution to the nucleon spin. The difference
between $h_1^q$ and $g_1^q$ provides important constraints to any model of the nucleon. Knowing the
quark tensor charges one can construct the isovector nucleon tensor charge $g_T = \delta u- \delta
d$, a fundamental property of the nucleon which, together with the vector and axial charges,
characterizes the nucleon as a whole. Since many years the tensor charge is being calculated with
steadily increasing accuracy by lattice QCD~\cite{Chen:2016utp}. More recently, its connection with
possible novel tensor interactions at the TeV scale in neutron and nuclear $\beta$-decays and its
possible contribution to the neutron electric dipole moment (EDM) have also been
investigated~\cite{Bhattacharya:2016zcn}, and possible constraints on new physics beyond the
standard model have also been derived \cite{Courtoy:2015haa}.

A good knowledge of the tensor charge requires measurements of similar accuracy for deuteron and
proton data. For this reason we have proposed~\cite{addendum2018} to CERN a new one-year (150 days)
measurement, scattering the M2 muon beam with 160 GeV/c momentum on a transversely polarised
deuteron target, as soon as the second long-shut down of the CERN accelerator complex (LS2) will be
over, using the COMPASS spectrometer in the configuration we have used it in 2010. In the future,
JLab12 will provide new and hopefully precise data on both the proton and the deuteron, but at
fairly large $x$-values, namely $x>0.05$. The main objective of our proposal is to measure from $x\simeq
0.3$ down to 0.003 and at larger $Q^2$ (up to $Q^2 \simeq 100$ (GeV/c)$^2$), thus our measurement
will be complementary to the JLab12 measurement and essential both to evaluate the tensor charges
and to access the PDFs of the sea quarks.
 
Our proposal has been accepted by CERN, and the experiment should run in 2021.

In the following, for reasons of space and time, I will detail only the case for transversity, but
it is clear that the new SIDIS asymmetry data, combined with the good precision HERMES and COMPASS
proton data, and with the future high precision JLab12 data, will allow also all the other $u$ and
$d$ distribution functions to be extracted with comparable accuracy. This is particularly important
for the Sivers distribution. The future EIC will possibly supersede the existing and the proposed
measurements, but the new COMPASS contribution will stay there for several years.

\section{Present and extrapolated uncertainties for
COMPASS transverse spin asymmetries}
\label{sseunodue}
\begin{figure}[tb]
\begin{center}
\includegraphics[width=0.8\textwidth]{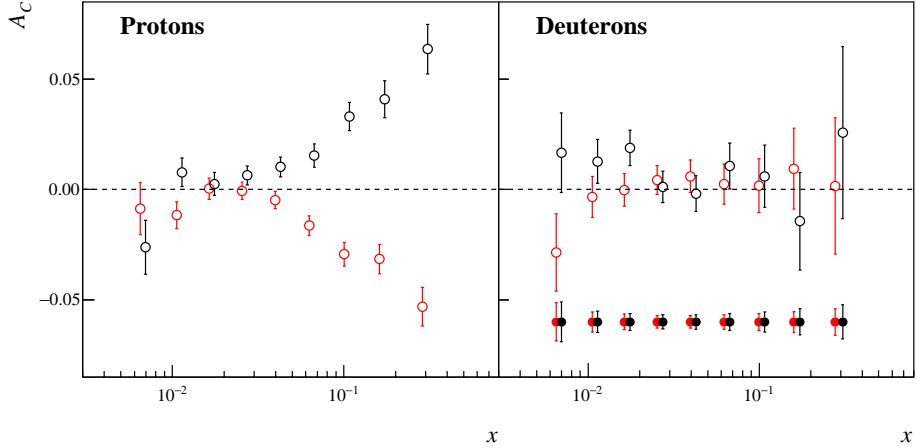}
\end{center}
\vspace{-0.5cm}
\caption{ The Collins asymmetry $A_{C}$ obtained from the 2010 data with the polarised proton NH$_3$
 target as a function of $x$ (left plot) compared to the results obtained from the runs of 2002,
 2003 and 2004 with polarised deuteron $^6$LiD target (right plot). The red (black) points refer
 to positive (negative) hadrons. The full points at $A_{C}=-0.06$ in the right plot show the
 extrapolated statistical errors from the proposed deuteron run.}
\label{fig:cpd}
\end{figure}

In order to estimate the impact of a future run on a transversely polarised deuteron target, it is
convenient to look at Fig.~\ref{fig:cpd}, where the Collins asymmetries for positive and negative
hadrons obtained from the 2010 data~\cite{Adolph:2012sn} collected using NH$_3$ as a polarised
proton target, are shown as a function of $x$ (left panel) and compared to the results obtained from
the deuteron runs of 2002, 2003, and 2004~\cite{Ageev:2006da}, when as target we used $^6$LiD (right
panel).
\begin{figure}[tb]
\begin{center}
\includegraphics[width=0.7\textwidth]{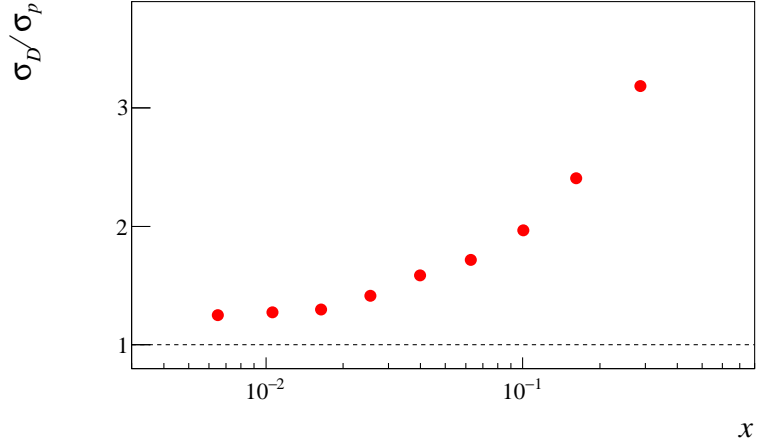}
\end{center}
\vspace{-0.5cm}
\caption{Ratio of the statistical uncertainties of deuteron and proton asymmetries for charged
 hadrons as measured by COMPASS.}
\label{fig:repd}
\end{figure}
Clearly the accuracy of the data is considerably better for the proton target. At small $x$ the
uncertainties are similar, but as a function of $x$, while the Collins asymmetries get larger, the
statistical uncertainties of the deuteron asymmetries become larger and larger than those of the
proton ones. As shown in Fig.~\ref{fig:repd}, where the ratio of the statistical uncertainties of
the asymmetries are plotted as function of $x$, the ratio is larger than a factor of three in the
last measured $x$-bin. The reason for this increase with $x$ is the fact that when the deuteron
target data were collected, due to the late delivery of the COMPASS polarised target (PT) magnet we
used the existing PT magnet previously constructed and used by the SMC Collaboration. The COMPASS PT
magnet was operational only in 2005, and we could use it for our transversely polarized proton runs
of 2007 and 2010. There is indeed a huge difference between the acceptance of the COMPASS PT magnet
and the SMC PT magnet: the COMPASS magnet has a polar angle acceptance of 180 mrad (as seen from the
upstream end of the target) while the SMC magnet has a corresponding acceptance of 70 mrad. A
reduced acceptance in scattering angle directly translates into a reduced acceptance at large
$x$-Bjorken.

By using the COMPASS PT magnet for the future deuteron run, in one year of data taking in the
conditions of the 2010 proton run we can safely expect the number of reconstructed SIDIS events to
be the same as in 2010, since target material densities and packing factors are almost identical for
$^6$LiD and NH$_3$. It is then very simple to estimate the statistical errors of the measured
Collins asymmetries in the future deuteron run, they will just be the statistical uncertainties of
the 2010 proton data scaled by the ratio of the figures of merit of the two targets, namely
\begin{eqnarray}
r=\frac{FOM_{p}}{FOM_{d}} = \frac{0.155 \cdot 0.80}{0.40 \cdot 0.50} = 0.62,
\label{eq:rat4}
\end{eqnarray}
where $FOM(=fP)$ is the figure of merit of the polarized target, $f$ is the dilution factor of the
target material, and $P$ is the proton or deuteron polarization. Under the assumption that the
spectrometer acceptance and efficiency be the same for the deuteron runs as they were for the 2010
proton run, the statistical errors of the transverse spin asymmetries will be smaller by a factor of
0.62 than those of the proton. The projected errors for the deuteron Collins asymmetries are shown
as closed points in the right plot of Fig.~\ref{fig:cpd}. We neglect the systematic errors which
were estimated to be at most 0.5 times the statistical errors in the 2010 data, so that they
increase the total error by less than 10\%.

Most important, very much as for the Collins asymmetry, all the target transverse spin asymmetries
(TSA) are expected to be measured with a statistical uncertainty equal to 0.62 times the statistical
uncertainties of the corresponding asymmetries measured in the 2010 proton run which in all the
following we use as a reference.

\section{Present and extrapolated uncertainties for transversity}  
\label{sec:aaa}
In order to quantify the gain in statistical uncertainty in $h_1^{u_v}$ and $h_1^{d_v}$, we have
extracted twice the $u$- and $d$-quark transversity, the first time using the existing COMPASS
deuteron data, and a second time using the projections of Fig.~\ref{fig:cpd} for the statistical
errors of the new deuteron data. In both cases all the existing proton data from HERMES and COMPASS
have been used. To carry through this evaluation we have followed the procedure of
Ref.~\cite{Martin:2014wua}, a point by point extraction of transversity directly from the measured
SIDIS and $e^+e^- \rightarrow hadrons$ asymmetries.

The big advantage of this method is that
\begin{itemize}
\item[-] it avoids the use of parametrizations for the unknown functions which introduce systematic
 errors difficult to estimate
\item[-] the transversity PDFs, their statistical uncertainties and their correlation coefficients
 can all be calculated algebraically from the measured asymmetries
\item[-] it allows to extract separately the valence quark transversity and the sea quark
 transversity as different linear combinations of the four measured spin asymmetries
 (positive/negative pions on proton/deuteron targets)
\end{itemize}but
\begin{itemize}
\item[-] it requires asymmetry measurements on different targets (p, d or n) available at the same
 $x$-values and in the same kinematic ranges.
\end{itemize}

\begin{figure}[tb]
\begin{center}
\includegraphics[width=0.45\textwidth]{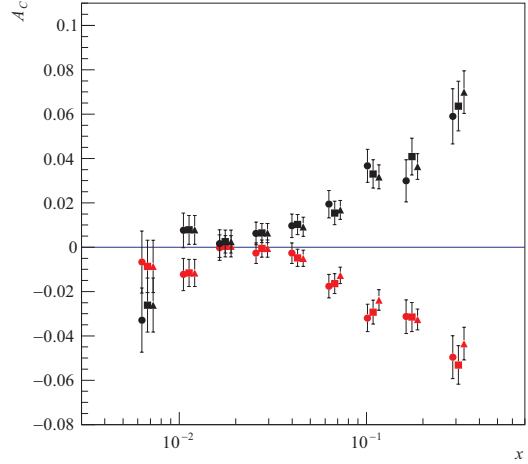}
\end{center}
\vspace{-0.5cm}
\caption{The Collins asymmetry for positive (red) and negative (black) hadrons
 from the existing proton data. In each $x$ bin, the first point (left to
 right) is from the 2010 COMPASS run, the second point is from the combined
 2007 and 2010 COMPASS data, the third is obtained by adding also the HERMES
 data.}
\label{fig:acollHC}
\end{figure}

For what concerns the last point, there has been no problem to add the results from the COMPASS
proton data collected in 2007~\cite{Alekseev:2010rw}: the data were taken with the same experimental
apparatus, at the same beam energy, have been binned in the same $x-$intervals, and we had already
merged them in the past and put the 2007 and 2010 combined results in HepData
\cite{Maguire:2017ypu}. However, merging the HERMES results with ours needs an approximation. For
$x>0.032$ the HERMES range overlaps with the COMPASS range, but the $x$ binning is
different~\cite{Airapetian:2010ds}. We have combined bins 2 and 3, and bins 5 and 6 of HERMES,
reducing the original 7 HERMES points into 5 points which are centered with good approximation in
our bins 5, 6, 7, 8, and 9. The HERMES data have also been corrected by $D_{NN}$ while the effect
of the cut $z<0.7$ and of the difference in energy have been neglected. A weighted mean of HERMES
and COMPASS data has then been done, and the transversity functions have been extracted point by
point as if all the data would have been taken by COMPASS. The overall situation is shown in
Fig.~\ref{fig:acollHC}.

\begin{figure}
\begin{center}
 \includegraphics[width=0.4\textwidth]{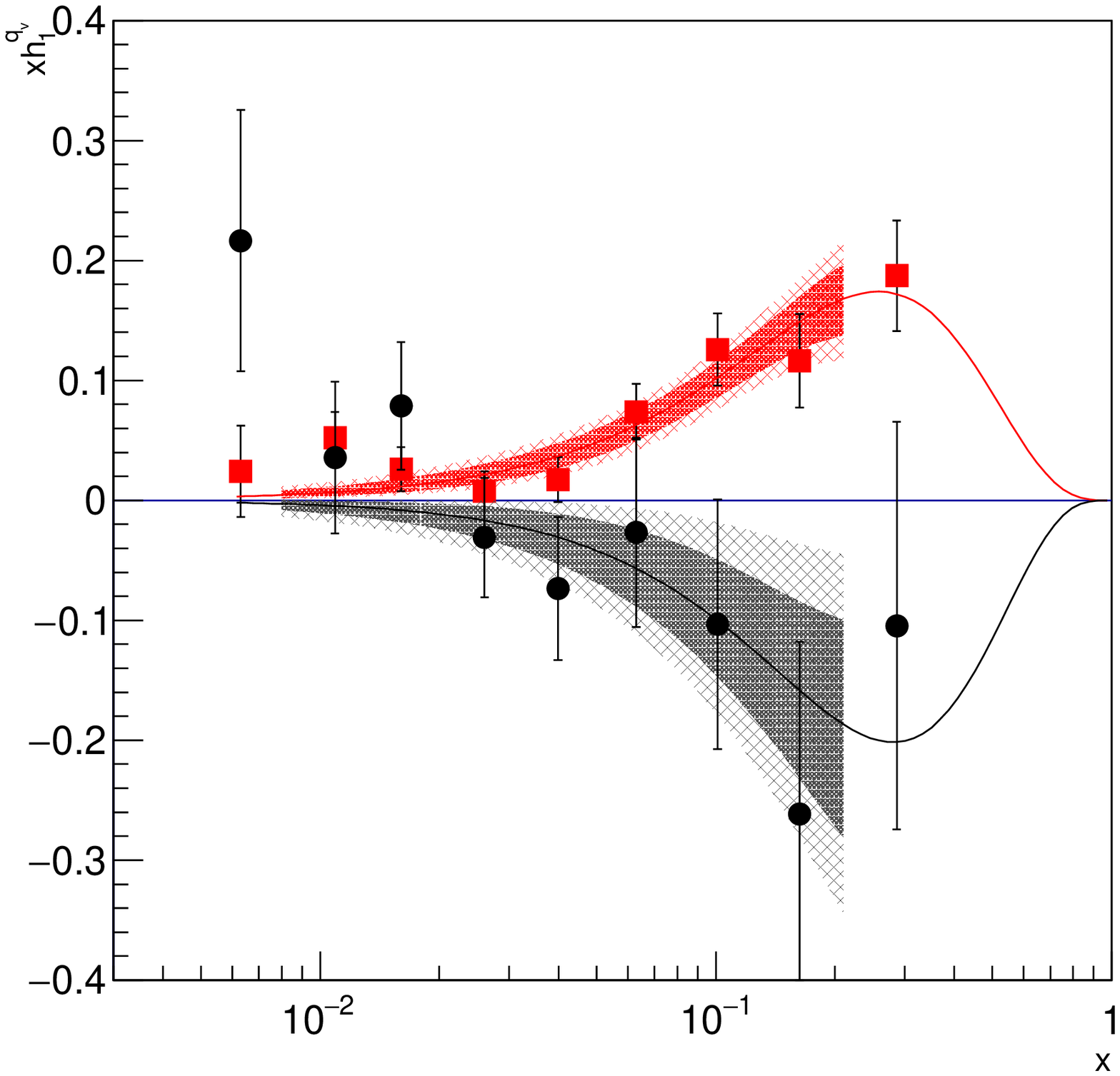}
 \hfill
 \includegraphics[width=0.4\textwidth]{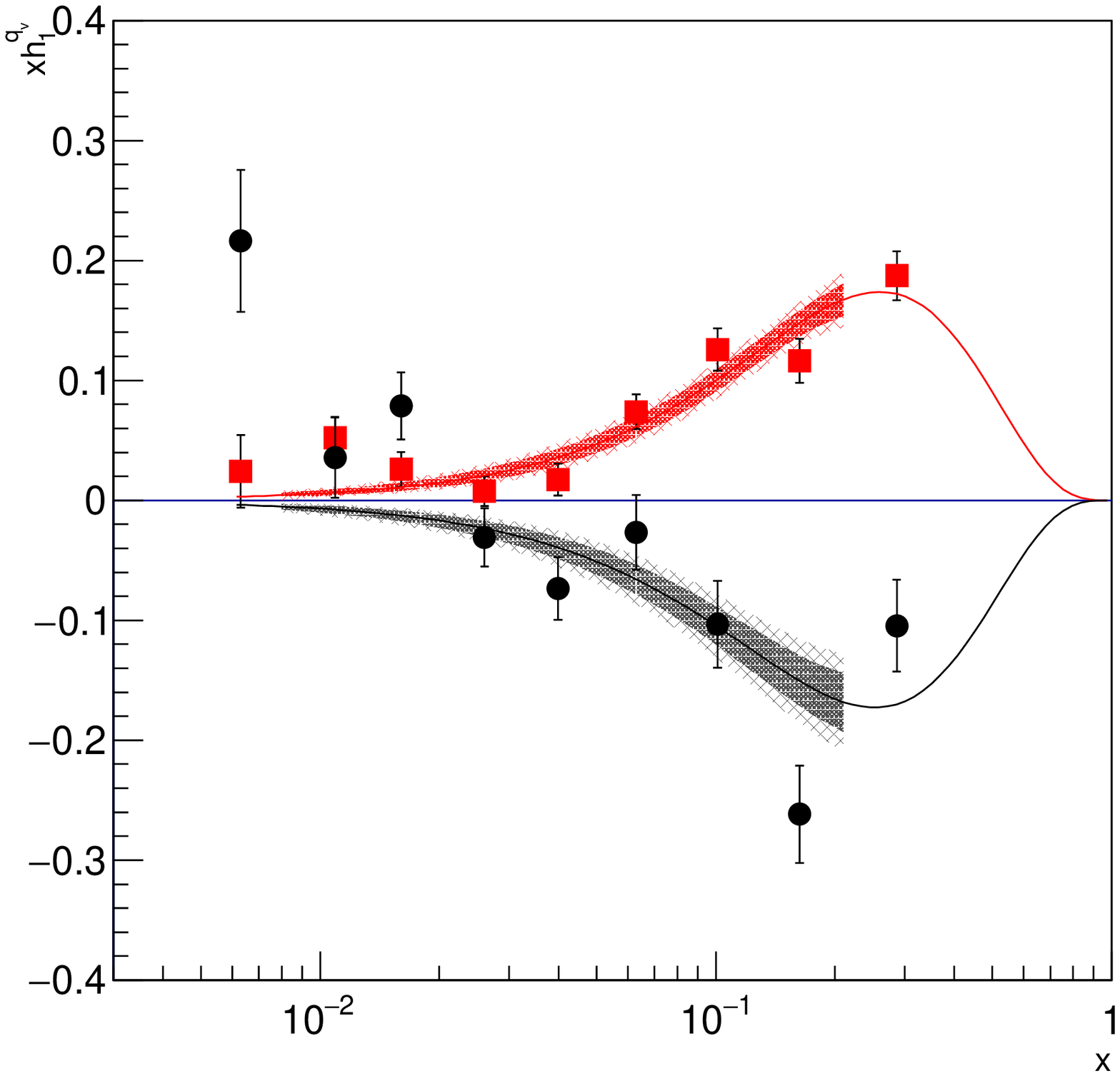}
\end{center}
\vspace{-0.5cm}
\caption{The points are the values for $xh_1^{u_v}$ (red squares) and $xh_1^{d_v}$ (black circles)
 extracted from the measured asymmetries. Left plot: existing deuteron data uncertainties. Right
 plot: projected uncertainties for the future deuteron run. Also shown, for both cases, are the
 68\% and 90\% confidence bands obtained from 80000 replicas.}
\label{fig12}
\end{figure}

The extracted values of $xh_1^{u_v}$ and $xh_1^{d_v}$ when using the 2021 projections for the
deuteron are shown in Fig.~\ref{fig12}, together with the 68\% and 90\% confidence bands obtained
from 80000 replicas (right panel), and compared with the corresponding quantities obtained with the
existing deuteron data (left panel).

As in Ref.~\cite{Martin:2014wua} we have used the Collins asymmetries for positive and negative
hadrons from the 2010 proton data and all the deuteron data assuming all the charged hadrons to be
pions, while identified pions are about 70\% of the ``all hadron'' sample. Using the identified
pion asymmetries, the statistical uncertainty would increase by about 20\%.

Since $xh_1^{u_v}$ and $xh_1^{d_v}$ are obtained as linear functions of the four measured
asymmetries (see Ref.~\cite{Martin:2014wua}) their estimated values are
correlated. Table~\ref{tab:ccCH} gives the correlation coefficients in the 9 $x$ bins, both for the
existing deuteron data and when using the projections of the proposed deuteron run. The correlation
coefficients strongly depend on the relative statistics between the proton and the deuteron data. In
the existing situation, which is largely unbalanced in favour of the protons, they are very close to
1, while they are close to zero with the new deuteron run.
\begin{table}
\begin{center}
\begin{tabular}{|l|r|r|}
\hline
$x$-bin & \multicolumn{2}{c|}{$\rho(xh_1^{u_v},xh_1^{d_v})$}\\
\hline
 & present & projected \\
\hline
0.003-0.008 & 0.494 & -0.029  \\
0.008-0.013 & 0.507 & -0.030  \\
0.013-0.020 & 0.509 & -0.042  \\
0.020-0.032 & 0.553 & -0.049  \\
0.032-0.050 & 0.641 & -0.009  \\
0.050-0.080 & 0.743 & 0.099  \\
0.080-0.130 & 0.764 & 0.033  \\
0.130-0.210 & 0.867 & 0.142  \\
0.210-0.70 & 0.863 & -0.100  \\ 
 \hline
\end{tabular}
\end{center}
\caption{Correlation coefficient between $xh_1^{u_v}$ and $xh_1^{d_v}$ in the different $x$ bins
 from numerical calculations, using the HERMES data, the 2007 and 2010 COMPASS proton data and
 present and projected deuteron errors.}
\label{tab:ccCH}
\end{table}

\section{Projections for the tensor charge uncertainties}
\label{sseunoduetre}
To evaluate the tensor charge the transversity PDFs for the valence quarks have to be integrated
over $x$ from 0 to 1. To avoid any possible bias, we prefer not to use any parametric function for
the transversity PDFs, and simply integrate numerically the measured valence quark transversity
values obtained as described in the previous section. The statistical uncertainty is then evaluated
by adding up the errors of the measurements in the various $x$-bins.

It is clear that COMPASS can give an important and unique contribution in a limited $x$-range, and
the evaluation of the tensor charge ultimately will rely on the measurements of many experiments.
In particular the SoLID Collaboration~\cite{Ye:2016prn} at JLab12 is expected to contribute with very accurate
measurements in the valence region, for $0.1<x<0.7$, in the future.

COMPASS, on the other hand, can give an accurate estimate of the tensor charge in the range
$0.008<x<0.21$. For different reasons, the first ($0.003<x<0.008$) and the ninth ($0.21<x<0.7$)
$x$-bins give fairly large contributions to the uncertainty of the tensor charge, therefore in the
following we integrate the transversity PDFs over the central seven bins, skipping the first and the
last.

The values of $\delta u$, $\delta d$ and $g_T$ in the selected $x$ range are given in
Table~\ref{tab:meth12CH}, both for the present uncertainties and using the projected errors for the
deuteron. The correlation coefficients of Table \ref{tab:ccCH} are properly taken into account in
the calculation of the uncertainties of the tensor charges.
\begin{table}
\begin{center}
\begin{tabular}{|l|c|c|c|c|c|c|}
\hline
 & $\delta u=\int dx h_1^{u_v}(x)$ & $\delta d=\int dx h_1^{d_v}(x)$ & $g_T=\delta u - \delta d$ \\
\hline
present &  
0.201 $\pm$ 0.032 & -0.189 $\pm$ 0.108 & 0.390 $\pm$ 0.087 \\ 
 projected & 
0.201 $\pm$ 0.019 & -0.189 $\pm$ 0.040 & 0.390 $\pm$ 0.044 \\
\hline
\end{tabular}
\end{center}
\caption{Truncated tensor charges from all existing proton and deuteron data
 (``present'') and with the projected uncertainties for the deuteron data
 (``projected''). The integration runs over the range $0.008<x<0.21$.}
\label{tab:meth12CH}
\end{table}
The reduction of the uncertainties on the integral of the $d$-quark when the projected measurements
are taken into account is considerable, from 0.108 to 0.040. This is the most important outcome we
expect from the new run we propose. However, also the knowledge of the $u$-quark transversity
integral improves, from 0.032 to 0.019. For the tensor charge $g_T$ the uncertainty goes from 0.087
to 0.044. The gain is smaller than for the $d$-quark case, for two reasons. First, the weights of
the $u$ and $d$ quarks in the tensor charge are the same, thus the gain for the tensor charge is a
sort of mean of the gains of the two PDFs. Secondly, with the present data the statistics of the
proton data is much larger than that of the deuteron data, and, as a consequence, the correlation
coefficients between the $u$- and the $d$- transversity is large and positive, making the variance
of the tensor charge smaller than the quadratic sum of the variances of the $\delta u$ and $\delta
d$. On the contrary, the correlation coefficient between $xh_1^{u_v}$ and $xh_1^{d_v}$ becomes
small (and negative) when one adds the new deuteron run, making the statistics of proton and
deuteron asymmetries similar, and the variance of the tensor charge is essentially the quadratic sum
of the variances of $\delta u$ and $\delta d$. It has to be stressed however that a situation in
which the extracted values of $xh_1^{u_v}$ and $xh_1^{d_v}$ are almost uncorrelated is preferable to
the one when $xh_1^{d_v}$ is strongly correlated to $xh_1^{u_v}$, which is the case presently.

The impact of the proposed deuteron measurement on the statistical accuracy of the tensor charge has
been evaluated also with two other methods, namely
\begin{enumerate}
\item
fitting with ``reasonable'' functions the extracted values of $xh_1^{u_v}(x)$ 
and $x h_1^{d_v}(x)$, and then integrating the curves obtained in
this way, and 
\item
using ``replicas'' of the measured asymmetries.
\end{enumerate}

The results obtained with these two more methods have slightly smaller uncertainties, but we prefer
to quote the ones obtained with the numerical integration which are fully unbiased.

To summarize, at large $x$, JLab12 will provide very accurate partial measurements for $g_T$. At
smaller $x$ ($0.008 < x < 0.21$) the COMPASS data will provide a contribution to $g_T$ with an
uncertainty of $\pm 0.044$. Without the new deuteron data from COMPASS, the evaluation of the
tensor charge from only the future high precision JLab data would be affected by the error of the
integration between $0$ and $0.1$, which is difficult to be ascertained, and the result will anyhow
be model dependent. We expect that, with the new COMPASS deuteron data, the uncertainty of the
extrapolated contribution of the integration from 0 to 0.003 will be much smaller, and the
uncertainty of the partial integration of the COMPASS data will be the resulting uncertainty on the
tensor charge $g_T$.

To summarize, the new COMPASS measurement on a transversely polarized deuteron target will allow to
reduce the uncertainties on the d quark PDFs. This will be particularly important for the
transversity PDF, which allows to access the tensor charge $g_T$. At small $x$ values ($0.008 < x <
0.21$) the COMPASS data will provide a contribution to the measurement of $g_T$. with an
uncertainty  of $\pm 0.044$. At large $x$,  JLab12 should provide  very  accurate  partial  measurements
for $g_T$, but without  the  new  deuteron  data  from  COMPASS  the  evaluation  of  the tensor
 charge would  be  affected  by  the  error  of the integration between 0 and 0.1, which is
difficult to be ascertained, and the result will anyhow be model dependent.  We expect that, with
the new COMPASS deuteron data, the uncertainty of the  extrapolated  contribution  of  the
 integration over $x$ from  0  to  0.003  will  be  much  smaller, and  the uncertainty of the
partial integration of the COMPASS data might be the resulting overall
uncertainty on the tensor charge $g_T$.

% Bibliography
%\bibliographystyle{elsart-num}
\bibliographystyle{PoS}
{
\raggedright
\bibliography{spin18fb}
}

\end{document}